\documentstyle[12pt,psfig]{ioplppt}

\begin{document}
\jl{1}
\bibliographystyle{plain}

\title{Low temperature phase diagram and critical behaviour of the four-state 
chiral clock model}[Four-state chiral clock model]

\author{M Pleimling\dag, B Neubert\ddag ~ and R Siems\ddag}
\address{\dag\ Institut f\"ur Theoretische Physik B, Technische Hochschule,
D--52056 Aachen, Germany, e-mail: pleim@physik.rwth-aachen.de}
\address{\ddag\ Fachrichtung Theoretische Physik, Universit\"at des Saarlandes, Postfach 151150, D--66041 Saarbr\"ucken, Germany}

\begin{abstract}
The low temperature behaviour of the four-state chiral clock ($CC_4$) model is reexamined
using a systematic low temperature series expansion of the free energy.
Previously obtained results for the low temperature phases
are corrected and the low temperature phase diagram is
derived. 
In addition, the phase transition from the modulated region to the high temperature paraphase
is shown to belong to the universality class of the 3d-$XY$ model.
\end{abstract}
\maketitle
%----------------------------------------------------------------------------------------------------
\newcommand{\frak}{\sf}
\renewcommand{\vec}{\bf}
\newcommand{\altvec}[1]{{\underline{\rm #1}}}
\newcommand{\altmat}[1]{{\underline{\underline{\rm #1}}}}
\renewcommand{\imath}{{\rm i}}
\newcommand{\emath}{{\rm e}}
\newcommand{\op}{\cal}
\newcommand{\func}{\rm}
\newcommand{\set}{\frak}
\newcommand{\dbidx}[3]{{#1}({#2},{#3})}
\newcommand{\vardbidx}[3]{{#1}({#2};{#3})}
\newcommand{\tridx}[4]{{#1}({#2};{#3},{#4})}
\newcommand{\quadidx}[5]{ {{#1} {\binom{{#2},{#3}}{{#4},{#5}}}} }
\newcommand{\sixidx}[7]{{#1} {\binom{{#2};{#3},{#4}}{{#5};{#6},{#7}}} }
\newcommand{\stackindex}[2]{{\stackrel{\scriptstyle #1}{\scriptstyle #2}}}
\newcommand{\avg}[1]{{\bigl< #1 \bigr>}}
\newcommand{\greenf}[2]{{\bigl<\bigl< #1;#2 \bigr>\bigr>}}
\newcommand{\chemical}[1]{{$\rm #1$}}
\newcommand{\transpose}[1]{{{#1}^{\rm T}}}
\newcommand{\phase}[1]{$\bigl < #1 \bigr >$}
\newcommand{\sign}{{\rm sign~}}
\newcommand{\trace}{{\rm tr~}}
%----------------------------------------------------------------------------------------------------
\noindent
\section{Introduction}
Uniaxially modulated structures are observed in very different
classes of magnetic and ferroelectric substances. In many cases they exhibit
rather complex phase diagrams with large varieties of phases. Phase 
transitions from a high temperature paramagnetic or paraelectric phase 
(paraphase) to commensurately and incommensurately modulated phases occur, as 
external control parameters like temperature 
and elastic stresses are varied. Microscopic models are successfully
used for the description of these modulated systems. 
They were reviewed e.g.\ in \cite{Sel88}. An interesting example is 
the $p$-state chiral clock model \cite{Yeo82}, whose Hamiltonian is
%\begin{eqnarray}
%H & = & - J_0 \sum\limits_{\alpha} \sum\limits_{<ij>} \cos \left[ \frac{2 \pi}{p}
%\left( n_{i, \alpha} - n_{j, \alpha} \right) \right] \nonumber \\
%& & - J \sum\limits_i \sum\limits_{\alpha} \cos \left[ \frac{2 \pi}{p}
%\left( n_{i, \alpha} - n_{i+1, \alpha} + \Delta
%\right) \right].
%\label{hamiltonian}
%\end{eqnarray}
\begin{equation}
\fl H = - J_0 \sum\limits_{\alpha} \sum\limits_{<ij>} \cos \left[ \frac{2 \pi}{p}
\left( n_{i, \alpha} - n_{j, \alpha} \right) \right] 
- J \sum\limits_i \sum\limits_{\alpha} \cos \left[ \frac{2 \pi}{p}
\left( n_{i, \alpha} - n_{i, \alpha+1} + \Delta
\right) \right].
\label{hamiltonian}
\end{equation}
$\alpha$ labels the layers perpendicular to the direction of the modulation
(chiral direction) and $i$, $j$ the crystal units in these layers.
$\left< ij \right>$ runs over neighbouring pairs in the layers.
The integer variables $n_{i, \alpha}$ describe the state of the unit
$(i,\alpha)$. They assume one of the values from
0 to $p-1$. Below they are called spins.
The two terms in equation (\ref{hamiltonian}) describe couplings ($J_0 > 0$,
$J > 0$)
between nearest neighbours in the same and in
adjacent layers, respectively.\\
In the ground state every layer is ferromagnetically ordered.
Depending on the value of $\Delta$, various ordering patterns of the different layers are 
realized. For $0 \leq \Delta < \frac{1}{2}$ nearest neighbours in the chiral direction
couple ferromagnetically (ferromagnetic bond),
thus leading to a ferromagnetic ground state
where all spins are equal. For $\frac{1}{2} 
< \Delta \leq 1$ 
the spin increases by one for successive layers (chiral bond),
thus yielding the right-handed chiral pattern
\begin{displaymath}
\ldots ~ 0 ~ 1 ~ 2 ~ \ldots ~ (p-1) ~ 0 ~ 1 ~ \ldots
\end{displaymath}
$\Delta = \frac{1}{2}$ is a multiphase point at which an infinity of different phases are
degenerate since ferromagnetic and chiral bonds have the same energy.\\
Whereas the three-state model ($p = 3$) \cite{Ost81,Hus81}
has been very thoroughly investigated, 
only few results are known for the general case $p \geq 4$. 
There are derivations of the low
temperature phase diagram of the general $p$-state model by an expansion of the free
energy in the vicinity of the multiphase point \cite{Yeo82} as well as by a low
temperature mean-field theory \cite{Yeo84}, in which it was claimed that, for
the four-state model ($p = 4$), only the phases $\left< 1 2^k \right>$, $\left< 1 2^k 1 2^{k+1}
\right>$, $\left< 2^k 3 \right>$, $\left< 2^k 3 2^{k+1} 3 \right>$, $\left< 4
\right>$, and $\left< \infty \right>$ ($k = 0, 1, 2, \ldots$) are stable at low temperatures.
$\left<  u_1 \ldots u_r \right>$ is a shorthand notation for the phase with a period
consisting of $r$ bands with $u_1, u_2, \ldots , u_r$ 
layers with spins $n$, $n+1, \ldots, n+r$ (all modulo $p$) respectively.
The phase $\left< 12 \right>$, for example, is given by the layer sequence
\begin{displaymath}
\ldots ~ 0 ~ 1 ~ 1 ~ 2 ~ 3 ~ 3 ~ 0 ~ 1 ~ 1 ~ \ldots
\end{displaymath}
The ferromagnetic and chiral ground states are denoted by $\left< \infty \right>$ 
and $\left< 1 \right>$ respectively.\\
McCullough \cite{Mcc92} investigated the phase diagram for
$p = 3$, 4, and 5 using the mean-field transfer-matrix (MFTM) method.
From the numerical 
extrapolation of the data it was concluded that the low temperature phase diagrams
for $p = 3$ and $p = 4$ were consistent 
with the results of the low-temperature
series expansion \cite{Yeo82,Yeo81}. It is interesting, that, for $p = 5$,
new phases not predicted by the
low-temperature series expansion \cite{Yeo82} were found to be stable at low
temperatures.\\
Scholten and King \cite{Sch96} presented Monte Carlo simulations of
the four- and the six-state models. They investigated especially the transition
from the modulated phases to the ferromagnetic phase (i.e. $\Delta < \frac{1}{2}$).
As it was not possible to resolve particular phases, they determined the "interface
spacing" as the average number of layers in a band for a given phase.
They claimed that, for $\Delta = 0.45$, the results were not
inconsistent with the predictions of Yeomans. In the case $p =4$ and $\Delta = 0.2$
new phases with an interface spacing larger than the interface spacings of the phases
predicted in \cite{Yeo82} were observed 
close to the transition to the ferromagnetic phase.\\
Recently the four-state chiral clock model was shown \cite{Ple97} to be 
a special case of the Double Ising Spin (DIS) model \cite{Ple94,Neu94,Ple96},
which was introduced to describe uniaxially modulated ferroelectrics.\\
In the following new results for the four-state chiral clock model are presented. 
In section \ref{sec2} and \ref{sec3}
we will reexamine the low temperature phase diagram and 
discuss discrepancies with previous results.
In section \ref{sec4} it is shown that the 
transition from the modulated phases to the paramagnetic phase belongs to the
universality class of the 3d-$XY$ model, and in section 5
short conclusions are given.

\section{The low temperature series expansion\label{sec2}}
The present series expansion technique for the four-state chiral clock ($CC_4$) model
is similar to the method
developed by Fisher and Selke \cite{Fis80} for the Axial Next Nearest 
Neighbour Ising (ANNNI) model. At low temperatures the reduced free energy
per spin $f = \frac{F}{Nk_BT}$ ($N$ is the total number of spins)
may be expanded in the form \cite{Fis80}
\begin{equation}
f = \frac{E_0}{k_BT} - \frac{1}{N} \sum\limits_{n \geq 1} \Delta Z_N^{(n)}.
\label{free_energy}
\end{equation}
$\Delta Z_N^{(n)}$ is the total contribution 
to the partition function from configurations in which 
$n$ spins have flipped (as compared to the ground state). $E_0$,
the ground state energy per spin, can be expressed \cite{Yeo82,Yeo81} in terms
of the structural variables \cite{Fis80}
$l_k = L_k/L$ ($L_k$: number of $k$-layer bands; $L$: total number of layers):
\begin{displaymath}
E_0 \left( \left\{ l_k \right\} \right) = - \frac{1}{2} q_\perp J_0 - J_1 
-J_1 \, \delta \, \sum\limits_{k \geq 1} l_k
\end{displaymath}
with $J_1 = J \cos \left( \frac{\pi}{2} \Delta \right)$ and $\delta = \tan \left( 
\frac{\pi}{2} \Delta \right) - 1$. The number of nearest neighbours in the 
layers is $q_\perp$; it is 4 for the primitive cubic lattice.\\
The contributions $\Delta Z_N^{(n)}$ are expressed
in terms of the elementary Boltzmann factors
\begin{displaymath}
\fl w = \exp \left( - K_0 \right), ~ x = \exp \left( - 2 K \, \cos \left( 
\frac{\pi}{2} \Delta \right) \right)
~~ \mbox{and} ~~ y = x^{1+ \delta} = \exp \left( - 2 K \, \sin 
\left( \frac{\pi}{2} \Delta \right) \right)
\end{displaymath}
with $K_0 = J_0/(k_BT)$ and $K = J/(k_BT)$. 
The reduced free energy per spin can be expanded in a convergent power series
of $w$, provided that $x \gg w$, i.e.\ if $J_0$ is large compared to $J$ (which
is assumed throughout this paper). 
The weight $w$ results from changing
an in-layer bond between spins with equal values to a bond between spins
with values differing by 1. 
The lowest orders involved are $w^{q_\perp}$ (overturning one spin),
$w^{2 q_\perp-2}$ (overturning two neighbouring spins in one layer) and
$w^{2 q_\perp}$ (overturning two spins not being in-layer nearest neighbours).\\
There are
three possible environments of a given spin (the numbers in
parentheses are the values of the spins in three consecutive layers where the
considered spin belongs to the middle layer): (a) spins with
two ferromagnetic bonds in the chiral direction (e.g.\ $0\hat{0}0$), (b) spins
with one ferromagnetic and one chiral bond (e.g.\ $0\hat{1}1$), and (c) spins 
with two chiral bonds (e.g.\ $0\hat{1}2$).\\
Let us discuss, as an example, the contribution
to $\Delta Z_N^{(1)}$ [first-order term in equation (\ref{free_energy})] for case (a). 
By overturning one spin, three different final states can be obtained
($m$ being the initial state): $(m+1) ~ \mbox{mod} ~ 4$, $(m+2) ~ \mbox{mod} ~ 4$, and 
$(m+3) ~ \mbox{mod} ~ 4$. This leads to the Boltzmann factor
\begin{eqnarray}
\fl \sum\limits_{n = 1}^3 \exp \left[ - \left( E_f(n) - E_i \right)/\left( k_BT \right) \right]
\nonumber \\
\lo=  \sum\limits_{n = 1}^3 \left( \exp \left\{ 2K \, \cos \left( \frac{\pi}{2} \Delta \right)
\, \left[ \cos \left( \frac{\pi}{2} n \right) -1 \right] \right\} \right. \nonumber \\
\left. \times
\exp \left\{ q_\perp \, K_0 \left[ \cos \left( \frac{\pi}{2} n \right) -1 \right] \right\} \right)
\nonumber \\
\lo= x \, w^{q_\perp} + x^2 \, w^{2q_\perp} + x \, w^{q_\perp}.
\label{boltz}
\end{eqnarray}
It is obvious from
equation (\ref{boltz}) that the process $m \longrightarrow (m+2) ~ \mbox{mod} ~ 4$ does not contribute
to the lowest order term in the expansion, as it has the same in-layer Boltzmann factor
$w^{2q_\perp}$ as the higher order process by which the values of two uncoupled
spins change by 1. In fact, this process of the order $w^{2q_\perp}$
does not even contribute to the 
lowest order correction term, which is of the order $w^{2q_\perp-2}$ (flipping of two
neighbouring spins in one layer \cite{Fis80}).\\
In reference \cite{Yeo82} the following contribution to $\Delta Z_N^{(1)}$ for the case (a) is
given:
\begin{eqnarray}
\fl \sum\limits_{n = 1}^3 \exp \left[ - \left( E_f(n) - E_i \right)/\left( k_BT \right) \right]
\nonumber \\
\lo= \sum\limits_{n = 1}^3 \exp \left\{ 2K \, \cos \left( \frac{\pi}{2} \Delta
\right) \, \left[ \cos \left( \frac{\pi}{2} n \right) -1 \right] \right\} \omega^{q_\perp}
\nonumber \\
\lo= \left( x + x^2 + x \right) \omega^{q_\perp}
\label{boltzyeo}
\end{eqnarray}
with
\begin{equation}
\omega = \sum\limits_{n = 1}^3 \exp \left\{ K_0 \left[ \cos \left( \frac{\pi}{2} n \right)
-1 \right] \right\} 
= w + w^2 + w .
\label{boltzyeo2}
\end{equation}
A comparison of equations (\ref{boltzyeo}) and (\ref{boltzyeo2}) 
with equation (\ref{boltz}) reveals that
the treatment of the in-layer bonds is erroneous in reference \cite{Yeo82}. 
The free energy is written in reference \cite{Yeo82} as an expansion in terms of the (erroneous)
Boltzmann factor $\omega$.
As a consequence, contributions from different
orders of the expansion are treated in reference \cite{Yeo82} as 
if they were of the same order. Thus, in our example,
the term $x^2$, resulting from the process $m \longrightarrow (m+2) ~ \mbox{mod}
~ 4$ and contributing to a higher order correction in the polynominal
expansion in $w$ [see equation (\ref{boltz})], contributes 
to the lowest order in \cite{Yeo82} [see equation (\ref{boltzyeo})]. This error
is repeated for all considered spin configurations and for all considered $p$-state
models ($p \geq 4$), thus
leading to a wrong low temperature phase diagram not only for the $CC_4$ model, but
also for the generalised $p$-state chiral clock model with $p \geq 4$. One should
emphasise that the treatment
of the in-layer bonds is correct in the analyses of the $CC_3$ model \cite{Yeo81}.\\
With the correct contributions, the reduced free energy (2) in first order is given by
\begin{equation}
\fl f = - \frac{1}{2} q_\perp K_0 - K_1 - \frac{1}{2} K_1 \delta - \left( 1 + x \, y \right) 
w^{q_\perp} + a_1( \delta ) \, l_1 + \sum\limits_{k \geq 3} a_k( \delta ) \, l_k
+ O(w^{2 q_\perp -2})
\label{f_firstorder}
\end{equation}
with 
\begin{displaymath}
a_1( \delta ) = - \frac{1}{2} K_1 \delta - \left( 2 y - x y -1 \right) w^{q_\perp}
\end{displaymath}
and
\begin{displaymath} 
a_k( \delta ) = \left( k - 2  \right) \left[ \frac{1}{2} K_1 \delta - \left( 2 x - x y -1   
\right) w^{q_\perp} \right].
\end{displaymath}
The set of structural variables $l_k$ minimizing
$f$ for given values of $\delta$ and $T$ determine the stable phases occurring in
first order (see figure 1): the $\left< \infty \right>$-, the $\left< 1 \right>$-,
and the $\left< 2 \right>$-phase.
Phases $\left< \infty \right>$ and $\left< 2 \right>$ are, in this order
of the expansion, separated by a boundary, at which all phases that are degenerate at
the multiphase point and that do not contain 1-layer bands have the same free energy. Likewise,
phases containing only 1- and 2-layer bands are still degenerate on the boundary
between the $\left< 1 \right>$- and the $\left< 2 \right>$-phase.\\
One could now proceed in considering processes involving two spins, then three
spins and so on. This is very cumbersome and only feasible for processes involving
few spins. 
In the next section the phases stable in general order in the series expansion
will be determined using a transfer-matrix method.

\section{Transfer-matrix method\label{sec3}}
\subsection{Introductory remarks}
One should first note that the Hamiltonian (\ref{hamiltonian}) is left invariant by
the transformation
\begin{eqnarray}
&& \Delta \longrightarrow \Delta ' = 1 - \Delta \nonumber \\
&& n_{i,\alpha} \longrightarrow n_{i,\alpha}' = \left( - n_{i,\alpha}
 + \alpha \right) ~ \mbox{mod} ~ 4 .
\label{trafo}
\end{eqnarray}
Therefore, the phase diagram of the $CC_4$ model is invariant under a reflection 
in the line $\Delta = \frac{1}{2}$. In the following we will discuss the low
temperature phase diagram for the case $\Delta > \frac{1}{2}$, i.e.\ we will analyse
in detail the stability of the boundary line between the $\left< 1 \right>$- and
the $\left< 2 \right>$-phase, the phase diagram for $\Delta < \frac{1}{2}$ being
inferred by the transformation (\ref{trafo}).\\
In the ground state and in the low temperature expansion every phase $\left< \nu
\right>$ consists of a periodic arrangement of a sequence of $n(\nu)$ layers called
$\nu$-sequences [$n(\nu)$ is the period of the phase]. Suppose now that in a certain
order of the series expansion two stable phases, $\left< \nu_1 \right>$ and 
$\left< \nu_2 \right>$, are separated by a boundary at which the phases produced by 
$\nu_1$- and $\nu_2$-sequences are degenerate (see figure \ref{fig2}).
In first order the boundary under
consideration separates the phases $\left< 1 \right>$ and $\left< 2 \right>$.
At higher order a new phase $\left< \nu \right> = \left< \nu_1 \nu_2 \right>$ 
consisting of a structure
with alternating $\nu_1$- and $\nu_2$-sequences might be stable in the vicinity
of the boundary. If 
\begin{equation}
a_\nu = f_{\left< \nu \right> } - \frac{n(\nu_1)}{n(\nu_1)+n(\nu_2)} 
f_{\left< \nu_1 \right> } - \frac{n(\nu_2)}{n(\nu_1)+n(\nu_2)}
f_{\left< \nu_2 \right> }
\label{anu}
\end{equation}
is negative, the new phase has a lower free energy than the phases $\left< \nu_1 \right>$ and 
$\left< \nu_2 \right>$ \cite{Fis80,Sen93} and it will be stabilized in the vicinity of the
$\left< \nu_1 \right>  :  \left< \nu_2 \right>$ boundary (see figure 2a).
The stability of the boundaries between the phases $\left< \nu_1 \right>$ and 
$\left< \nu_1 \nu_2 \right>$ and the phases $\left< \nu_1 \nu_2 \right>$ and
$\left< \nu_2 \right>$ must then be examined at higher orders.
If, on the other hand, $a_\nu$ is positive,
the phase $\left< \nu_1 \nu_2 \right>$ (and therefore every phase consisting of
$\nu_1$- and $\nu_2$-sequences) has a higher free energy than either $\left< \nu_1 \right>$
or $\left< \nu_2 \right>$. The boundary is a true phase boundary which remains stable
in all orders of the low temperature series expansion (see figure 2b).\\
The reader is referred to references \cite{Fis80} and \cite{Yeo81}
for details concerning
the construction of the series expansion to general order.\\

\subsection{Formulation in terms of transfer matrices and vectors\label{sec3b}} 
The sign of $a_\nu$, and therefore the stability of the phase $\left< \nu \right>$,
is determined by the leading term in its expansion in terms of $w$.
This term is obtained by considering all flipping processes involving
a spin chain of $n(\nu)-1$ spins in $n(\nu)-1$ different layers
\cite{Yeo81}. Besides the linear configuration with all $n(\nu)-1$ spins 
connected, the various decompositions of this configuration into 2, 3, $\ldots$, $n(\nu)-1$
different parts must be taken into account.
The contributions from these processes can be written
as a product of transfer matrices and vectors. 
The matrices describe a bond between two flipping spins, the vectors 
an initial or a final bond preceding the first or following the last flipped spin
respectively.\\
Every spin can flip to three different values and hence
$3 \times 3$ matrices occur. As we are only interested in the
sign of the $a_\nu$ we can restrict ourselves to the two processes contributing
in lowest order, thus excluding the process $m \longrightarrow (m+2) ~ \mbox{mod}
~ 4$ only relevant for the correction term. Of course,
if one considers all possible processes (i.e.\ $3 \times 3$ matrices),
the leading term is identical to the term obtained by the $2 \times 2$ matrices.
This has already been noticed in the low temperature analyses of a six-state
clock model with competing axial nearest and next-nearest neighbour couplings \cite{Sen93},
where the corresponding $2 \times 2$ matrices have been considered instead
of the general $5 \times 5$ matrices.\\
As two axial next nearest neighbours
are either coupled by a ferromagnetic or by a chiral bond, only two different matrices
are to be constructed. For a ferromagnetic or a chiral bond between two spins
in the layers $\alpha$ and $\alpha$+1 one obtains, respectively, the transfer matrices
\begin{equation}
\altmat{F}_{\alpha, \alpha+1} =
\left( \begin{array}{ll} 
1-x ~ & ~ x(1-y) \\
x(1-y^{-1}) ~ & ~ 1-x
\end{array} \right) w^{q_\perp}
\label{matrix_ferro}
\end{equation}
and
\begin{equation}
\altmat{C}_{\alpha, \alpha+1} =
\left( \begin{array}{ll}
1-y ~ & ~ y(1-x^{-1}) \\
y(1-x) ~ & ~ 1-y
\end{array} \right) w^{q_\perp}.
\label{matrix_chiral}
\end{equation}
The matrix elements are the Boltzmann factors
for a simultaneous change of the values of the two
spins. The first (second) row corresponds to a
change $\Delta n_{i,\alpha} = +1 (-1)$ and
the first (second) column to $\Delta n_{i,\alpha+1} = +1 (-1)$.
Every element of the matrices \altmat{F} and \altmat{C} is a sum of two terms,
the first term resulting from changing the values of two axially coupled 
spins. As already mentioned, disconnected pairs of spins
(i.e.\ two spins that are not neighbour to each other but neighbour
to an unchanged spin) also contribute to the partition sum. Since every
disconnected pair must be associated with a minus sign \cite{Fis80}, the
corresponding Boltzmann factors enter the different matrices
with a negative sign.\\
The factor $w^{q_\perp}$ resulting from changing 
the in-layer bonds in layer $\alpha$+1
is common to all elements of the matrices \altmat{F} and \altmat{C}.
This is a direct consequence of the fact that only flipping processes
$m \longrightarrow (m \pm 1) ~ \mbox{mod} ~ 4 $ are to be considered for
obtaining the leading order in the expansion of $a_\nu$.
For the full $3 \times 3$ matrices this is not
the case as the flipping process $m \longrightarrow (m+2) ~ \mbox{mod} ~ 4 $ has the in-layer
Boltzmann factor $w^{2 q_\perp}$.
In reference \cite{Yeo82} the phase diagram has been determined to general order using $3 \times 3$
transfer matrices. Due to the erroneous treatment of the in-layer interactions 
(see section \ref{sec2}) the "common term" $\omega^{q_\perp}$ has been
factorized, thus leading, again,
to the treatment of terms belonging to different orders as being of the same order.\\
A spin at the end of the spin chain is neighbour of an unchanged spin. To determine
the contributions of these spins, four different cases are to be distinguished:
(a) the considered spin is the first spin of the chain and its bond to the left (i.e.\
to an unchanged spin) is a ferromagnetic or a chiral bond (subscripts $f$ and $c$
respectively) or (b) it
is the last spin of the chain and its bond to the right is 
a ferromagnetic or a chiral bond. The Boltzmann factors for the
flipping of these single spins are written as vectors:
\begin{eqnarray}
\altvec{a}_f & = & \left(
\begin{array}{l}
y^{- \frac{1}{2}} \\
y^{\frac{1}{2}}
\end{array} \right) \, x^{\frac{1}{2}} \, w^{q_\perp} 
\label{af} \\
\altvec{a}_c & = & \left(
\begin{array}{l}
x^{\frac{1}{2}} \\
x^{- \frac{1}{2}}
\end{array} \right) \, y^{\frac{1}{2}} \, w^{q_\perp} 
\label{ac} \\
\altvec{b}_f & = & \left(
\begin{array}{l}
y^{\frac{1}{2}} \\
y^{- \frac{1}{2}}
\end{array} \right) \, x^{\frac{1}{2}} 
\label{bf} \\
\altvec{b}_c & = & \left(
\begin{array}{l}
x^{- \frac{1}{2}} \\
x^{\frac{1}{2}}
\end{array} \right) \, y^{\frac{1}{2}}
\label{bc}
\end{eqnarray}
The vectors (\ref{bf}) and (\ref{bc}) do not include the Boltzmann factor resulting from 
the change of the in-layer bonds. This factor has already been included in the matrix
describing the overturning of the two last spins in the spin chain.

\subsection{Derivation of the low temperature phase diagram}
With the matrices (\ref{matrix_ferro}) and (\ref{matrix_chiral}) and the vectors
(\ref{af})-(\ref{bc}) it is now possible to compute the leading order term $b_\nu$
of the quantities $a_\nu$ (and, thus, to determine the sign of $a_\nu$) for all phases
degenerate at the multiphase point and containing only 1- and 2-layer bands.
All considered phases can be viewed as periodic arrangements
of spin sequences with a 1-layer band as the first and a 2-layer band as the last band
in the sequence 
\cite{Yeo81}. The sequence $\tilde{\nu}$ obtained by stripping
the original sequence $\nu$ by its last and first band is called core. All sequences
based on the same core $\tilde{\nu}$ enter in the computation of the $b_\nu$: The
sequences $ 1 \tilde{\nu} 2$ and $ 2 \tilde{\nu} 1$ contribute negatively, the
sequences $ 1 \tilde{\nu} 1$ and $ 2 \tilde{\nu} 2$ contribute positively \cite{Fis80}.
The expressions $b_\nu$ for different families of phases 
are summarised in table 1.

\subsubsection{Stability of some series of phases}
For the series of phases $\left< 1 2^k \right>$ the expression
(see table 1)
\begin{equation}
b_{12^k} = - \left( \altvec{a}_c^T - \altvec{a}_f^T \right) \left( \altmat{C} \, 
\altmat{F} \right)^{k-1} \altmat{C} \left( \altvec{b}_f - \altvec{b}_c \right)
\label{b12k}
\end{equation}
gives the leading contribution to $a_{12^k}$. 
The four different sequences based on the core $\tilde{\nu}= 2^{k-1}$ yield
the four different contributions to $b_{12^k}$.\\
The eigenvalues $\exp \left( - \Gamma_\pm \right)$ of the 
matrix $\altmat{C} \, \altmat{F}$ are 
real and positive. Expression (\ref{b12k}) can be written in the form
\begin{displaymath}
b_{12^k} = A_+ \, \exp \left( - \frac{k}{2} \Gamma_+ \right) + A_- \, \exp \left( - \frac{k}{2}
\Gamma_- \right)
\end{displaymath}
with $\Gamma_+ < \Gamma_-$. A close
examination reveals that for finite temperatures $A_+ < 0$, $A_- > 0$ and $A_+ + A_- < 0$.
Thus, $b_{12^k}$ is negative for all $k$, i.e.\ all phases of the form
$\left< 1 2^k \right>$ spring from the multiphase point and have a finite stability 
range at temperatures above zero.\\
The leading order contribution for the phases $\left< 1^k 2 \right>$ is
\begin{equation}
b_{1^k2} = - \left( \altvec{a}_c^T - \altvec{a}_f^T \right) \altmat{C}^k 
\left( \altvec{b}_f - \altvec{b}_c \right).
\label{b1k2}
\end{equation}
The eigenvalues of the matrix $\altmat{C}$ are complex conjugate. They are
written in the form
\begin{equation}
\xi_1 \pm \i \xi_2 = \exp \left( - \Gamma_0 \pm \i \, \Omega \right)
\label{eigen}
\end{equation}
with $\Gamma_0 = - \frac{1}{2} \ln \left( \xi_1^2 + \xi_2^2 \right) > 0$, 
$\Omega = \arctan \left( \frac{\xi_2}{\xi_1} \right)$ and $\xi_1 = 1 -x^{1+\delta}$, $\xi_2 =
\left( 1 - x \right) x^{\frac{1}{2} + \delta}$. We then obtain the expression
\begin{displaymath}
\fl
b_{1^k2} = - (A_1 + \i A_2) (\xi_1 + \i \xi_2 ) + (A_1 - \i A_2) (\xi_1 - \i \xi_2 )
= - \left| \Delta \right| \exp \left( - k \, \Gamma_0 \right) \cos \left(
k \, \Omega + \phi \right)
\end{displaymath}
with $\left| \Delta \right| \exp \i \phi = A_1 + \i A_2$.
The temperature-dependent quantities $\left| \Delta \right|$, $\phi$, 
$\Gamma_0$, and $\Omega$ do not depend on $k$.\\ 
$b_{1^k2}$ is negative for small
values of $k$. If $k$ exceeds the value $k_{max} = \frac{1}{\Omega} \left( \frac{\pi}{2} -
\phi \right)$, then $b_{1^k2}$ becomes positive and, thus, all phases with
$k > k_{max}$ are unstable at the considered point of the phase diagram. 
Since $k_{max} \longrightarrow \infty$ for $T \longrightarrow 0$, there is, 
for every $k$, a temperature
below which the phase $\left< 1^k 2 \right>$ is stable. Thus, all phases 
$\left< 1^k 2 \right>$ spring from the multiphase point, but the higher
commensurate phases disappear at higher temperatures. Such a cut-off of the
high commensurate phases at finite temperatures is also observed in the ANNNI 
model \cite{Fis87}.\\
Following the general line we
also examined the series of phases $\left< 1 2^k 1 2^{k+1} \right>$ and
$\left< 1^k 2 1^{k-1} 2 \right>$.
For the case $\left< 1 2^k 1 2^{k+1} \right>$ we find that all these phases are
stable at finite temperatures in the vicinity of the multiphase point with no
cut-off for the phases with a large value of $k$, i.e.\ the results for the series
$\left< 1 2^k 1 2^{k+1} \right>$  resemble the results for the series $\left< 1 2^k \right>$.
Analysing the leading contribution for the phases $\left< 1^k 2 1^{k-1} 2 \right>$
we find a behaviour similar to the behaviour of the phases $\left< 1^k 2 \right>$,
i.e.\ all phases with $k < k_{max}$ (the value of $k_{max}$ being series-dependent)
are stable and $k_{max} \longrightarrow \infty$ as $T \longrightarrow 0$.

\subsubsection{Phases containing general sequences of 1- and 2-layer bands}
In the following we will show
that all phases consisting only of 1- and 2-layer bands and obeying the rules
of the structure combination
spring from the multiphase point, the higher commensurate phases of some series
becoming unstable at higher temperatures.
The leading contribution to $a_\nu$ for all these phases is
of the form (see table 1)
\begin{equation}
b_\nu = - \left( \altvec{a}_c^T - \altvec{a}_f^T \right) \altmat{D} \, 
\altmat{C} \left( \altvec{b}_f - \altvec{b}_c \right)
\label{bnu}
\end{equation}
where $\altmat{D}$ is a product of powers of matrices $\altmat{C}$ 
and $\left( \altmat{C} \, \altmat{F} \right)$. The contributions of the
first and last band are given by $\left( \altvec{a}_c^T - \altvec{a}_f^T \right)$
and $\altmat{C}
\left( \altvec{b}_f - \altvec{b}_c \right)$ respectively.
A 1-layer band in the core contributes a
matrix $\altmat{C}$, whereas a 2-layer band yields the matrix product
$\left( \altmat{C} \, \altmat{F} \right)$. The product over all bands
in the core yields the matrix $\altmat{D}$ [see equation (\ref{bnu})].\\
The diagonal elements of the matrix
\begin{equation}
\fl 
\altmat{C} \, \altmat{F} = \left( \begin{array}{ll}
2 \left( 1 -y \right) \left( 1 - x \right) & x \left( 1 - y \right)^2 - x^{-1} y
\left( 1 - x \right)^2 \\
y \left( 1 - x \right)^2 - x y^{-1} \left( 1 - y \right)^2 & \left( 1 
+ x y \right) \left( 1 -y \right) \left( 1 - x \right)
\end{array} \right) \, w^{2 q_\perp}
\nonumber
\end{equation}
are positive whereas the non-diagonal elements are negative, since
$y = x^{1 + \delta}$ with $x \ll 1$. We now follow reference
\cite{Sen93} and introduce the unitary matrix
\begin{displaymath} 
\altmat{U} = \left( \begin{array}{ll}
-1 &  ~0 \\
~0 & ~1 
\end{array} \right) = \altmat{U}^{-1}.
\end{displaymath} 
All elements of $\altmat{U} \, \left( \altmat{C} \, \altmat{F} \right) \, \altmat{U}$,
and therefore of
$\altmat{U} \, \left( \altmat{C} \, \altmat{F} \right)^k \, \altmat{U}$, are positive.
This is also the case for
the two vectors $\left( \altvec{a}^T_c - \altvec{a}^T_f \right) \, \altmat{U}$
and $\altmat{U} \, \altmat{C} \, \left( \altvec{b}_f - \altvec{b}_c \right)$,
thus [see equation (\ref{b12k}) and table 1]
\begin{displaymath}
\fl \left( \altvec{a}_c^T - \altvec{a}_f^T \right) \left( \altmat{C} \, 
\altmat{F} \right)^{k-1} \altmat{C} \left( \altvec{b}_f - \altvec{b}_c \right)
= \left( \altvec{a}_c^T - \altvec{a}_f^T \right) \, \altmat{U} \, \altmat{U} \,
\left( \altmat{C} \, \altmat{F} \right)^{k-1} \, \altmat{U} \, \altmat{U} \,
\altmat{C} \left( \altvec{b}_f - \altvec{b}_c \right)
\end{displaymath}
is positive, i.e.\ $b_{12^k} < 0$, in agreement
with the aforementioned calculations.\\
Phases of the series $\left< 1 2^k 1 2^{k+1} \right>$ contain a single
1-layer-band in the core yielding the matrix product $\left( \altmat{C} \, \altmat{F} 
\right) \altmat{C} \left( \altmat{C} \, \altmat{F} \right)$ with positive diagonal
and negative non-diagonal elements for small $x$. Hence, the product (see table 1)
\begin{eqnarray}
\fl \left( \altvec{a}_c^T - \altvec{a}_f^T \right) \left( \altmat{C} \,
\altmat{F} \right)^k \altmat{C} \left( \altmat{C} \,
\altmat{F} \right)^k \altmat{C} \left( \altvec{b}_f - \altvec{b}_c \right)
\nonumber \\
\lo= \left( \altvec{a}_c^T - \altvec{a}_f^T \right) \, \altmat{U} \, \altmat{U} \,
\left( \altmat{C} \, \altmat{F} \right)^{k-1} \, \altmat{U} \, \altmat{U} \,
\left( \altmat{C} \, \altmat{F} \right) 
\altmat{C} \left( \altmat{C} \, \altmat{F} \right) \, \altmat{U} \nonumber \\
\times \altmat{U} \,
\left( \altmat{C} \, \altmat{F} \right)^{k-1} \altmat{U} \, \altmat{U} \,
\altmat{C} \left( \altvec{b}_f - \altvec{b}_c \right)
\nonumber
\end{eqnarray}
is positive, showing the stability of the phases $\left< 1 2^k 1 2^{k+1} \right>$.
Following this line of thought one easily shows that all phases appearing between 
the phases $\left< 2 \right>$ and $\left< 12 \right>$ (i.e. phases with only isolated
1-layer-bands in the core) are stable in the vicinity of the multiphase point.
Indeed, as no new matrix products show up in the computation of 
the different $b_\nu$, all these
expressions can be written, using the matrix $\altmat{U}$,
as a product of vectors and matrices having only positive elements.\\
%Furthermore, all allowed phases appearing
%between the phases $\left<  12 \right>$ and $\left< 2 \right>$ spring from the 
%multiphase point and are stable at finite temperatures (i.e. there is no cut-off
%of high commensurate phases), as only isolated 1-layer bands do appear, thus leading to
%matrices and vectors with only positive elements . In reference
%2 only the existence of the phases $\left< 1 2^k \right>$ and $\left< 1 2^k 1 2^{k+1}
%\right>$ was shown.\\
For phases containing consecutive 1-layer-bands in the core
the following additional vectors and matrices may contribute 
to the $b_\nu$ as can be seen
from table 1: $\altmat{U} \, \altmat{C}^k \, \left( \altvec{b}_f - \altvec{b}_c \right)$,
$\left( \altvec{a}_c^T - \altvec{a}_f^T \right) \, \altmat{C}^k \,
\altmat{U}$, and $\altmat{U} \, \left( \altmat{C} \, \altmat{F} \right) \,
\altmat{C}^k \, \altmat{U}$ with $k \geq 2$. Introducing
the eigenvalues of the matrix $\altmat{C}$ [see equation (\ref{eigen})], we obtain
\begin{eqnarray}
\mbox{v}_1 & = & \exp \left( -k \Gamma_0 \right) \left[ x^\frac{\delta}{2} \left( 1 - x \right)
\cos k \Omega + x^{-\frac{1+\delta}{2}} \left( 1 - x^{1+\delta} \right)
\sin k \Omega \right] \, w^{k q_\perp} \nonumber \\
\mbox{v}_2 & = & \exp \left( -k \Gamma_0 \right) \left[ x^{-\frac{\delta}{2}} 
\left( 1 - x^{1+\delta} \right) \cos k \Omega - x^{1+\delta} \left( 1 - x \right)
\sin k \Omega \right] \, w^{k q_\perp}  \nonumber
\end{eqnarray}
for the components of the vector $\altvec{v}= \altmat{U} \, \altmat{C}^k \,
\left( \altvec{b}_f - \altvec{b}_c \right)$.
Whereas $\mbox{v}_1$ is always positive, $\mbox{v}_2 > 0$ only if the
inequality 
\begin{displaymath}
\tan k \Omega < \frac{x^{-\frac{\delta}{2}} \, \left( 1 - x^{1+ \delta } \right)}
{x^{1+ \delta } \, \left( 1 - x \right)}
\end{displaymath}
holds. This is the case for temperatures smaller than an upper limit which
depends on $\delta$ and $k$.
In a similar way one shows that for temperatures smaller than some $k$-dependent
temperature all the components of the vector 
$\left( \altvec{a}_c^T - \altvec{a}_f^T \right) \, \altmat{C}^k \,
\altmat{U}$ and of the matrix $\altmat{U} \, \left( \altmat{C} \, \altmat{F} \right) \,
\altmat{C}^k \, \altmat{U}$ are positive.
The free energy differences $a_\nu$ for all phases 
containing consecutive 1-layer-bands in the core are therefore negative below
a certain temperature, 
i.e.\ these phases possess a stability region below this
temperature.

\subsubsection{Conclusion}
The results obtained so far can be summarised as follows: All
phases consisting only of 1- and 2-layer-bands, that can be formed
by means of the aforementioned structure combination rules, spring from
the multiphase point, where they are degenerate. The higher commensurate
phases of some series, i.e.\ those phases formed in higher orders of the combination
process, disappear again at temperatures individually depending
on the series under consideration.\\ 
From these results the complete low temperature phase diagram
of the $CC_4$ model is deduced by applying the transformation (\ref{trafo}).
At non-zero temperatures all phases appearing between
the phases $\left< 12 \right>$ and $\left< 3 \right>$ are stable since the 
transformation (\ref{trafo}) transforms phase $\left< 12 \right>$ into
$\left< 3 \right>$ and leaves the phase $\left< 2 \right>$ invariant. 
Some of the long commensurate phases
appearing between the phases $\left< 1 \right>$ and $\left< 12 \right>$
for $\Delta > \frac{1}{2}$ 
and between the phases $\left< \infty \right>$ and $\left< 3 \right>$
for $\Delta < \frac{1}{2}$ are unstable at a given temperature.
Upon reducing the temperature,
more and more of these phases become stable, and in the limit $T \longrightarrow 0$ all phases
obeying the rules of the structure combination
are stable. Therefore, the $CC_4$ model
exhibits a complete devil's staircase in the low-temperature limit.

\subsection{Comparison with other work}
The low-temperature behaviour of the general $p$-state chiral clock model was
analysed in reference \cite{Yeo82} using a series expansion technique similar 
to the one presented here. Due to the incorrect expansion (see section 2)
only some specific families of phases were shown to 
possess a finite stability region at small temperatures. Especially, it was
claimed that the phases $\left< 1^k 2 \right>$ with $k > 2$ are not stable
at low temperatures, implying, due to the transformation (\ref{trafo}), that
for $\Delta < \frac{1}{2}$ a direct transition from 
the ferromagnetic phase to the $\left< 4 \right>$-phase exists.
In order to corroborate these calculations a low temperature mean-field analyses 
of the $CC_p$ model was presented in \cite{Yeo84} were it was claimed that
in the vicinity of the multiphase point the mean-field approximation
yields the same stable phases as reference \cite{Yeo82}. In that work the model in mean-field
approximation was mapped onto an one-dimensional array of interacting domain walls. This
mapping was derived under the approximation that 
the mean-field average spin
$( \langle \cos \frac{\pi}{2} n_{i,\alpha} \rangle_{MF},
\langle \sin \frac{\pi}{2} n_{i,\alpha} \rangle_{MF} )$ 
in each layer (layer spin)
does only deviate from
the $T = 0$ value in amplitude but not in phase.
In a detailed analyses of the mean-field phase diagram of the $CC_3$ model 
Siegert and Everts \cite{Sie85} showed that
this approximation leads to a wrong phase diagram at low temperatures, thus concluding
that the layer spin must also be allowed to deviate in phase from its ground-state
value. This should not only be the case for the three-state but also for the
general $p$-state model. The results of reference \cite{Yeo84} for the 
mean-field low temperature
behaviour of the $CC_p$ model must therefore be considered with care.\\
As we have shown in the preceding sections the results of the series
expansion in reference \cite{Yeo82} are erroneous due
to wrong Boltzmann factors for the in-layer bonds. In fact, the four-state
model exhibits in the low-temperature limit a complete devil's staircase.
Furthermore, it results from our calculations that no direct
transition from the ferromagnetic to the $\left< 4 \right>$-phase exists
as phases with longer periods are stable between these two phases.\\
In the Monte Carlo simulation of the $CC_4$
model \cite{Sch96} long-period spin patterns were observed
when going from the ferromagnetic phase to the modulated phases
at rather high temperatures.
In view of the present work one must interpret
these patterns as reflecting the existence of phases
springing from the multiphase point and intercalating
between the ferromagnetic
and the $\left< 4 \right>$-phases.

\section{The critical behaviour\label{sec4}}
The critical behaviour of the general $p$-state chiral clock 
model at the transition to the paraphase is an interesting topic since 
for $p=2$ the chiral clock model reduces to the anisotropic Ising model, for
$p = \infty$ it corresponds to the classical 3d-$XY$ model. Siegert and Everts \cite{Sie89}
showed that the $CC_3$ model belongs to the universality class of the 3d-$XY$ model.
On the basis of his contradictory MFTM results, McCullough \cite{Mcc92} speculated about
a change in the universality class from 3d-Ising behaviour to 3d-$XY$ behaviour for
$p$ close to 5.
In the following we will show that for $p=4$ an
effective Ginzburg-Landau-Wilson Hamiltonian can be derived which can be
transformed to the effective Hamiltonian of the 3d $XY-$model.\\
For the case $p=4$
the Hamiltonian (\ref{hamiltonian}) can be rewritten
in the form
\begin{displaymath}
H = -J \sum\limits_{i} \sum\limits_{\alpha} \altvec{S}_{i,\alpha} \, \altmat{R} \left( 
\Delta \right) \,
\altvec{S}_{i,\alpha+1} - J_0 \sum\limits_{\alpha} \sum\limits_{\left< ij \right>} 
\altvec{S}_{i,\alpha} \, \altvec{S}_{j,\alpha}
\end{displaymath}
where we introduced the spin vector $\altvec{S}_{i,\alpha} = \left( 
\cos \frac{\pi}{2} n_{i,\alpha},
\sin \frac{\pi}{2} n_{i,\alpha} \right)$ and the rotation matrix
\begin{displaymath}
\altmat{R} \left( \Delta \right) = \left( \begin{array}{ll}
~ \cos \frac{\pi}{2} \Delta & \sin \frac{\pi}{2} \Delta  \\
- \sin \frac{\pi}{2} \Delta & \cos \frac{\pi}{2} \Delta 
\end{array}
\right).
\end{displaymath}
Rotating all spins in layer $\alpha$ by the angle $\frac{\pi}{2} \alpha \Delta$, i.e.\
introducing new vectors $\altvec{\sigma}_{i,\alpha}= \altmat{R} \left( \alpha \Delta \right) \,
\altvec{S}_{i,\alpha}$, leads to the expression
\begin{equation}
Z=\sum\limits_{\left\{ \altvec{\sigma} \right\} } \exp \left[- \frac{1}{2} \sum\limits_{ij}
\sum\limits_{\alpha \, \beta} \sum\limits_{\kappa=1}^{2} \sigma_{i,\alpha}^\kappa \,
K_{i \, \alpha, j \, \beta} \, \sigma_{j, \beta}^\kappa \right]
\label{partition}
\end{equation} 
for the partition function, $\kappa$ labelling the two spin components. The elements 
$K_{i \, \alpha, j \, \beta}$ of the coupling matrix are zero unless the lattice sites
$\left( i,\alpha \right)$ and $\left( j , \beta \right)$ are nearest neighbours.\\
Expression (\ref{partition}) may be transformed \cite{Bak62,Hub72} to
\begin{displaymath}
\fl
Z = C \sum\limits_{\left\{ \altvec{\sigma} \right\}} \left( \prod_{\rho=1}^2 \prod_{k \, \gamma} 
\int dh_{k, \gamma}^\rho \right) \exp \left[ - \frac{1}{2} \sum\limits_{ij}
\sum\limits_{\alpha \, \beta} \sum\limits_{\kappa} h_{i,\alpha}^\kappa \,
L_{i \, \alpha, j \, \beta}^{-1} \, h_{j, \beta}^\kappa + 
\sum\limits_{i} \sum\limits_{\alpha} \sum\limits_{\kappa} h_{i,\alpha}^\kappa \,
\sigma_{i,\alpha}^\kappa \right].
\end{displaymath}
Here $C$ is a numerical constant, $N$ is the total number of lattice sites and $\altmat{I}$ is 
the $N \times N$ identity matrix. The matrix $\altmat{L}$ is given by
$\altmat{L}= \mu \, \altmat{I}- \altmat{K}$ where the positive number $\mu$ is
chosen large enough to ensure that all the eigenvalues of $\altmat{L}$ are positive.\\
The sum over all states can be easily computed:
\begin{eqnarray} 
\fl 
\sum\limits_{\left\{ \altvec{\sigma} \right\}}  \exp \left( 
\sum\limits_{i} \sum\limits_{\alpha} \sum\limits_{\kappa} h_{i,\alpha}^\kappa \,
\sigma_{i,\alpha}^\kappa \right) \nonumber \\
\lo= 2^N \prod_{i \, \alpha} \left[ \cosh \left(
h^1_{i, \alpha} \, c_\alpha - h^2_{i, \alpha} \, s_\alpha \right) + \cosh \left(
h^1_{i, \alpha} \, s_\alpha + h^2_{i, \alpha} \, c_\alpha \right) \right],
\label{partcosh}
\end{eqnarray}
with $c_\alpha = \cos \left( \frac{\pi}{2}
\alpha \Delta \right)$ and $s_\alpha = \sin \left( \frac{\pi}{2}
\alpha \Delta \right)$. Using the expansion 
\begin{displaymath}
\ln \cosh x = - \sum\limits_n \left( - 1 \right)^n
\frac{2^{2n-1} \left( 2^{2n} -1 \right) B_n}{n \left( 2 n \right) !} x^{2n}
\end{displaymath}
($B_n$: Bernoulli number) the expression on the righthand side of (\ref{partcosh})
can be written as the exponential of a sum of powers of $h_{i,\alpha}^1$ and
$h_{i,\alpha}^2$:
\begin{equation}
\fl
C_1 \, \exp \left[ \sum\limits_n \sum\limits_{k=0}^{2n} \sum\limits_{l=0}^k \sum\limits_{l'=0}^{2n-k}
c(n,k,l,l') \sum\limits_i \sum\limits_\alpha c_\alpha^{l+l'} \, s_\alpha^{2n-l-l'}
\, \left( h^1_{i, \alpha} \right)^k \, \left( h^2_{i, \alpha} \right)^{2n-k}
\right]
\label{h1h2exp}
\end{equation}
where $c(n,k,l,l')$ is a number depending on $n$, $k$, $l$ and $l'$.\\
Introducing new variables $\phi^\kappa_{i,\alpha} = 
L_{i \, \alpha, j \, \beta}^{-1} \, h_{j, \beta}^\kappa$, taking the continuum limit
and turning to the wavenumber representation leads to the 
partition function
\begin{displaymath}
Z \propto  \left( \prod\limits_{\altvec{q}} \int d\altvec{\tau}_{\altvec{q}} \right) \exp \left[ - \bar{H} \right]
\end{displaymath}
with the effective Hamiltonian
\begin{eqnarray}
\fl \bar{H} = - \frac{1}{2} \int_{BZ} \frac{d^3q}{( 2 \pi )^3} \left(
r + q^2_{\perp} + \frac{\Upsilon}{\Upsilon_0} q^2_{\parallel} \right) \left[
\underline{\tau} ( \underline{q} ) ~ \underline{\tau} ( - \underline{q} )
\right] \nonumber \\[2mm]
- u \int_{BZ} \frac{d^3q \, d^3q' \, d^3q''}{( 2 \pi )^9}
\left[ \underline{\tau} ( \underline{q} ) ~ \underline{\tau} (
\underline{q}' ) \right] \, \left[ \underline{\tau} ( \underline{q}'' )
~ \underline{\tau} (- \underline{q} - \underline{q}' - \underline{q}'' )
\right]
\label{effham}
\end{eqnarray}
with $\altvec{\tau}_{\altvec{q}}=\left( \tau_{\altvec{q}}^1 , \tau_{\altvec{q}}^2
\right)$ 
and $\phi^{\kappa} \left( \altvec{r} \right) = \int_{BZ} \frac{d^3q}{( 2 \pi )^3} \,
\exp ( \i \altvec{q} \cdot \altvec{r} ) \, \tau_{\altvec{q}}^\kappa$.
The integration is over the first Brillouin zone with $\altvec{q} = \left( 
\altvec{q}_{\perp}, \altvec{q}_{\parallel} \right)$, 
its components $\altvec{q}_{\perp}$ and $\altvec{q}_{\parallel}$ being perpendicular
and parallel to the direction of the modulation respectively.
$r = \frac{1}{\Upsilon_0} \left( 1 - 2 \Upsilon_0 - \Upsilon \right)$ with
$\Upsilon = \frac{J}{k_BT}$ and $\Upsilon_0 = \frac{J_0}{k_BT}$ 
varies linearly with temperature.\\
In deriving equation (\ref{effham}) we neglected
fourth and higher harmonics, i.e.\ fast oscillating terms containing
$\exp \left( \mbox{i} n \frac{\pi}{2} \alpha \Delta \right)$ with 
$n \geq 4$. Furthermore
we did not include terms of higher than fourth order in $\tau$. If we rescale
$q_\parallel$ in the effective Hamiltonian \cite{Nel75} we end with the
effective Ginzburg-Landau-Wilson Hamiltonian of the 3d-$XY$ model. 

\section{Conclusions\label{sec5}}
A low temperature series expansion technique is suitable to obtain exact
results on the low 
temperature behaviour of the four-state chiral clock model. 
All phases degenerate at the multiphase point ($T=0$, $\Delta=\frac{1}{2}$)
and obeying the structure combination rules spring from
the multiphase point. Some of these phases disappear at higher temperatures.
In the low-temperature limit the $CC_4$ model exhibits
a complete devil's staircase.
Differences in the low temperature phase diagrams derived in the present and in
a previous publication can be traced back to an inconsistency in the series
expansion of the latter.
Long-period spin patterns derived in the present paper as 
stable phases between the ferromagnetic and
the $\left< 4 \right>$-phase and not occurring in the analyses presented 
in \cite{Yeo82}, were recently
seen in Monte Carlo
simulations just above the boundary of the ferromagnetic phase.\\
Furthermore,
the critical behaviour at the boundary between the paraphase and 
the modulated structures follows from the derivation of
an effective Ginzburg-Landau-Wilson
Hamiltonian. It is shown that the latter can be 
transformed to the effective Hamiltonian of the
3d-$XY$ model. The four-state model thus belongs to the universality class
of the $XY$ model.

\Tables
\begin{table}
\caption{The leading orders $b_\nu$ in the expansion of 
the quantities $a_\nu$
determing the stability of different 
families of phases consisting of 1- and 2-layer-bands.}
\begin{indented}
\item[]\begin{tabular}{@{}ll}
\br
$\nu$ & $b_{\nu}$ \\
\mr
$1^k2$ & $\left( \underline{a}^T_c - \underline{a}^T_f \right)
\underline{\underline{C}}^k \left( \underline{b}_f -
\underline{b}_c \right)$ \\
$1^k21^{k-1}2$ & $\left( \underline{a}^T_c - \underline{a}^T_f \right)
\underline{\underline{C}}^k \, \underline{\underline{F}}
\, \underline{\underline{C}}^k \left( \underline{b}_f -
\underline{b}_c \right)$ \\
$12^k$ & $\left( \underline{a}^T_c - \underline{a}^T_f \right)
\left( \underline{\underline{C}} \, \underline{\underline{F}}
\right)^{k-1} \, \underline{\underline{C}} \left( \underline{b}_f -
\underline{b}_c \right)$ \\
$12^k12^{k+1}$ & $\left( \underline{a}^T_c - \underline{a}^T_f \right)
\left( \underline{\underline{C}} \, \underline{\underline{F}}
\right)^k \underline{\underline{C}}
\left( \underline{\underline{C}} \, \underline{\underline{F}}
\right)^k \underline{\underline{C}} \left( \underline{b}_f -
\underline{b}_c \right)$ \\
\br
\end{tabular}
\end{indented}
\end{table}

\Figures
\begin{figure}
\caption{Schematic phase diagram showing, for given small value of $T$,
the phase sequence on a line along which $\delta$ varies.
Lower horizontal line: sequence in zeroeth order
of the expansion (exact for $T=0$); upper line: first order results.
$\delta =0$ is the multiphase point. Dashed lines
indicate boundary lines at which an infinity of phases are degenerate.}
\label{fig1}
\end{figure}

\begin{figure}
\caption{Schematic phase diagram in the vicinity of the boundary between
two stable phases $\left< \nu_1 \right>$ and $\left< \nu_2 \right>$. 
The horizontal lines correspond, as in figure 1, to a given value of $T$ and
present results of $n$th and $m$th order.
(a) The
phase $\left< \nu_1 \nu_2 \right>$ is stable at higher order leading to new
boundary lines. (b) The phase $\left< \nu_1 \nu_2 \right>$ is not stabilized
leading to a true phase boundary between the two phases $\left< \nu_1 \right>$
and $\left< \nu_2 \right>$. A dashed line indicates a
boundary at which an infinity of phases are degenerate, a solid line
indicates a true phase boundary.}
\label{fig2}
\end{figure}

\section*{References}
\begin{thebibliography}{99}
\bibitem{Sel88} Selke W 1988 {\it Phys. Rep.} {\bf 170} 213
\nonum Neubert B, Pleimling M and Siems R, {\it Ferroelectrics} (in print)
\bibitem{Yeo82} Yeomans J M 1982 \JPC 
{\bf 15} 7305
\bibitem{Ost81} Ostlund S 1981 \PR B {\bf 24} 398
\bibitem{Hus81} Huse D A 1981 \PR B {\bf 24} 5180
\bibitem{Yeo84} Yeomans J M 1984 \JPC {\bf 17} 3601 
\bibitem{Mcc92} McCullough W S 1992 \PR B {\bf 46} 5084
\bibitem{Yeo81} Yeomans J M and Fisher M E 1981 \JPC
{\bf 14} L835 
\nonum Yeomans J M and Fisher M E 1984 {\it Physica}
{\bf 127A} 1 
\bibitem{Sch96} Scholten P D and King D R 1996 \PR B {\bf 53} 3359
\bibitem{Ple97} Pleimling M, Neubert B and Siems R 1997 \ZP B {\bf 104} 125
\bibitem{Ple94} Pleimling M and Siems R 1994 {\it Ferroelectrics} {\bf 151} 69
\bibitem{Neu94} Neubert B, Pleimling M, Tentrup T and Siems R 1994
{\it Ferroelectrics} {\bf 155} 359
\bibitem{Ple96} Pleimling M and Siems R 1996 {\it Ferroelectrics} {\bf 185} 103
\bibitem{Fis80} Fisher M E and Selke W 1980 \PRL {\bf 44} 1502
\nonum 
Fisher M E and Selke W 1981 \PTRS {\bf 302} 1
\bibitem{Sen93} Seno F, Rabson D A and Yeomans J M 1993 \JPA
{\bf 26} 4887
\bibitem{Fis87} Fisher M E and Szpilka A M 1987 \PR B {\bf 36} 5343
\bibitem{Sie85} Siegert M and Everts H U 1985 \ZP B {\bf 60} 265
\bibitem{Sie89} Siegert M and Everts H U 1989 \JPA {\bf 22} L783
\bibitem{Bak62} Baker G A 1962 \PR {\bf 126} 2071
\bibitem{Hub72} Hubbard J 1972 \PL {\bf 39A} 365
\bibitem{Nel75} Nelson D R and Fisher M E 1975 \PR B {\bf 11} 1030

\end{thebibliography}
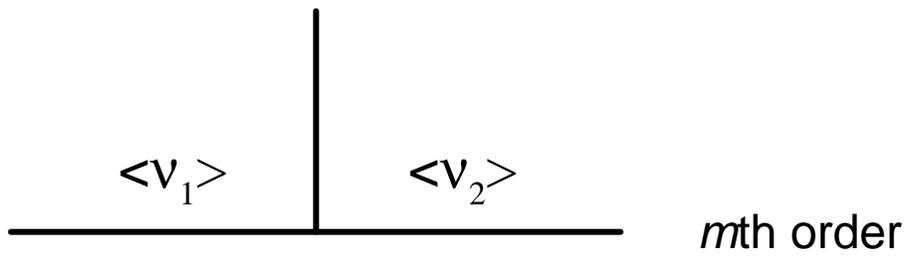
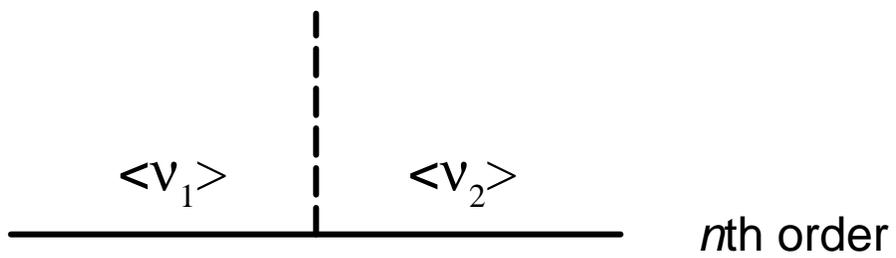
\end{document}